\documentclass[link, titlepage]{IWCOMP}

\copyrightyear{2013}
\usepackage{amssymb}
\usepackage{graphicx}
\usepackage{latexsym}
\usepackage{booktabs}
\usepackage{subfig}
\usepackage{epstopdf}
\usepackage[normalem]{ulem}
\usepackage{enumitem}
\usepackage{apalike}

\newcommand {\otoprule }{\midrule [\heavyrulewidth ]}

\usepackage{url}
\usepackage{theorem}
\usepackage{amsmath}
\usepackage[matrix,arrow]{xy}
\xyoption{curve}
\usepackage{textcomp}

\usepackage[usenames,dvipsnames,svgnames,table]{xcolor}

\newif\ifcomment

 \newcommand{\ignore}[1]{}

\DOI{xxxxxx}

\commentfalse

\begin{document}

\title[Should I Stay or Should I Go? Improving Event Recommendation in the Social Web]{Should I Stay or Should I Go?\\ Improving Event Recommendation\\ in the Social Web}

\author{Federica Cena,$^{1}$ Silvia Likavec,$^{1}$ Ilaria Lombardi,$^{1}$ and Claudia Picardi$^{1}$}

\affiliation{$^{1}$Dipartimento di Informatica, Universit\`a di Torino, Italy\\ Corso Svizzera 185, 10149 Torino, Italy}

\shortauthors{F. Cena, S. Likavec, I. Lombardi and C. Picardi}

\begin{abstract}

This paper focuses on the recommendation of events in the Social Web, and addresses the problem of finding if, and to which extent, certain features, which are peculiar to events, are relevant in predicting the users' interests and should thereby be taken into account in recommendation.

We consider in particular three ``additional'' features that are usually shown to users within social networking environments: reachability from the user location, the reputation of the event in the community, and the participation of the user's friends. Our study is aimed at evaluating whether adding this information to the description of the event type and topic, and including in the user profile the information on the relevance of these factors, can improve our capability to predict the user's interest.

We approached the problem by carrying out two surveys with users, who were asked to express 
their interest in a number of events. 
We then trained, by means of linear regression, a scoring function defined as a linear combination of the different factors, whose goal was to predict the user scores. We repeated this experiment under different hypotheses on the additional factors, in order to assess their relevance by comparing the predictive capabilities of the resulting functions.

The compared results of our experiments show that additional factors, if properly weighted, can improve the prediction accuracy with an error reduction of 4.1\%. The best results were obtained by combining content-based factors and additional factors in a proportion of approximately $10:4$. 
\end{abstract}


\category{recommender systems; contextual factors; user studies}

\maketitle

\section{Introduction}
\label{sec:introduction}

In recent years, recommender systems have been gaining popularity in e-services since they solve the problem of information overload by tailoring the system response to the preferences and needs of each specific user \citep{Ricci:11}. More precisely, recommender systems select and rank items (goods, news, services, web resources, etc.) so that users are presented only with potentially interesting information. Recommender systems extrapolate users' interests and preferences from different sources, from explicit user ratings to the implicit observation of user behavior \citep{Kobsa:00}. 

To some extent, we can say that recommender systems support users in making their choices, i.e.\ what to buy, what to look at, what to read, etc. It becomes then increasingly important that recommenders take into account as many factors as possible among those that affect users' choices~\citep{Jameson:11}, in order to be effective in their job. Many of these factors are actually circumstantial, and can be hardly captured by an automated system, whose knowledge of the user's context and whereabouts is necessarily limited. Nonetheless, the widespread availability of geo-positioning systems embedded in smartphones and mobile devices, the ubiquitous access to information and services guaranteed by cloud computing, the availability of social information recorded within the social web, and the resulting psychological willingness to publicly share private information, have widened significantly the amount and type of information that a recommender can exploit to its purposes. Then a legitimate question is: what type of information should a recommender use, and how, in order to provide a valid recommendation? Of course the answer partly depends on the nature of the objects to be recommended.

In this work we focus on the recommendation of {\em events}. Events are peculiar objects to recommend. As ``one-and-only items''~\citep{Cornelis:07a}, they usually happen in a specific time and place, have a limited availability, and are often experienced in groups. Often they need to be recommended before anyone has had the possibility to attend them (beside recurrent events) and feedback is available only for past events, which are not those to be recommended. By \emph{event} we mean anything with the following features: (a) having a participatory aspect, i.e.\ an activity is offered and people can decide whether to join or not, (b) requiring the physical presence of people at the location where the activity takes place, (c) being offered within a given time interval and requiring a certain amount of time to be completed. 

In this paper, we investigate some factors that, in the context of the social web, are known to users when they check out events 
and that may consequently influence their choices. These factors may include, for example, the presence of friends, the reputation or the popularity of the event (which is very often difficult to ascertain in advance in an automatic way), 
the distance 
to be covered in order to reach the event location, time constraints, etc. While the general problem of user choice (see e.g.\ \cite{citeulike:3835881,citeulike:2602869,citeulike:3128860,weber}) is out of the scope of this paper, our goal is to study the connection between such factors and the users' perceived interest in participating in an event, in order to provide empirical evidence on which information, and to which extent, should be taken into account by a recommender system for events.

The present lack of effective services that support users in finding and managing events in social applications suggests that event recommendation is not easily implemented by standard existing techniques. The most known and widespread example is Facebook\footnote{http://www.facebook.com}. The popular social networking environment offers the possibility to create events and make them known to other people, by either personal invitation or public sharing. Users can decide whether to join an event based on information such as a free-text description of its content, which of their friends have joined, where the event takes place (shown on a map) and of course its date and time.
However, there is no attempt at recommending events by ranking them according to any preferential order, or by basing them on their location (even when they take place thousands of miles away from the user position), or at including reputation mechanisms (events cannot be ``liked'' with the popular ``thumbs-up'' Facebook icon). At the time of this writing, the only type of suggestion provided by Facebook consists in showing public events that are connected to the user by the ``high-rank'' portion of her social graph.\footnote{The part of the social graph, including people, pages and places, that the user often checks out or comments on. This is the same criterion used by Facebook to choose which posts to show on the home page.} 

In order to provide a successful recommender system for this type of environments, we believe that a thorough analysis of the relevant factors is necessary and can contribute to the design and development of effective recommender systems. 

We assume that an event is first and foremost characterized by its content. In particular, we consider as content factors  the {\em themes} of the event (i.e.\ the topics the event is related to, as for example ``music'', ``rock'' and ``70's''), since they can reasonably represent the event topic, and the event {\em type} (i.e.\ the nature of the activity it involves, as for example ``concert'' or ``talk'').

We then focus on three additional factors that are easily available in a social networking environment in order to investigate the impact of their addition to content-related information on the accuracy of recommendation process:
\begin{description}
\item[Reachability] measures how feasible it is, in terms of distance, for the user to attend the event, taking into account the user's position and her willingness to move. Reachability can be regarded as a {\em spatial factor}, although it has a subjective component as it takes into account the users' personal preferences on the matter. 
\item[Friends' participation] is a {\em social factor} which takes into account how many among the users' friends have already expressed the intention to attend the event. 
\item[Average rating] expresses the overall community opinion on the upcoming event. It can be regarded also as a {\em social factor.} 
\end{description}

It is worth mentioning here that, although temporal properties (e.g.\ date, time, frame of availability, expected duration, etc.) are key parameters in recommendation of events, they are not part of our investigation. Adding the temporal dimension to our study would highly increase the variability of the answers and limit our capability to investigate the impact of the factors we were interested in. Instead, we regard temporal properties as a simple filter in the recommendation that rules out those events that are either impossible for the user to attend, or too far away in time to be worth recommending. 

The questions we aimed at answering with our study were the following:
\begin{itemize}
\item [RQ1] As users typically know more about an event than its themes and type, how does this knowledge affect their choices and the recommendation accuracy?
\item [RQ2] If we provide the system with additional information, does this enhance recommendation accuracy? 
\item [RQ3] To which extent should each additional factor contribute to the recommendation process? 
\item [RQ4] Would letting the users explicitly voice their preferences regarding additional factors provide any improvement in the recommendation? 
\item [RQ5] Does the relevance of additional factors depend on how much the user is interested in the event content? 
\end{itemize}

In order to answer these questions, we conducted two surveys in which we asked users to express their interest in participating in a certain event. In the first survey the same question was asked twice, initially disclosing only the theme and type of event, and subsequently showing additional information. In the second survey users immediately knew all the relevant information about the event. 

We then applied machine learning (ML) techniques to the collected data, in order to determine the relevance of the different factors in predicting the scores assigned by users to events. Machine learning is widely used in all the phases of recommender systems design: preliminary analyses (e.g.\ feature selection and construction, user profiling), offline setup (e.g.\ design and configuration of the recommendation algorithm), online update and maintenance (e.g.\ adaptation to user behavior, inference of user features), exploitation of results (e.g.\ presentation of recommendations, information extraction from user interaction.
See \cite{Adomavicius:05:surv} for a comprehensive survey on ML techniques in this field.

In our study we applied a predictive learning strategy: our goal at this stage was not to decide whether to recommend an event or not (a task for which a classificatory approach is generally more effective), but to assess the relevance of the different factors in predicting the user scores. We trained, using linear regression, a scoring function expressed as a linear combination of the factors we wished to analyze, with the aim of predicting the score assigned by a user to a given event. More precisely, we trained several scoring functions, each corresponding to different initial assumptions on which factors should be included and how they should be combined. 
This type of scoring function can be seen as a basic content-based recommender ~\citep{Pazzani:2007}, as each factor either matches the user profile w.r.t.\ the event features (e.g. how {\em reachable} the user perceives the event to be), or directly considers an event feature which is the same for all users (e.g. the event rating).

We turned the research questions above into hypotheses that could be tested by comparing the performance of the different scoring functions.

We used data collected in the first survey for training and validation, with a 15-fold random split of the samples in two groups (respectively, 2/3 of the samples for training and 1/3 of the samples for validation). We then tested the resulting functions on the data from the second survey, and compared their performance by means of paired t-test. 

While we refer the reader to the rest of the paper for a detailed discussion of the experimental results, we briefly anticipate here our core findings. Comparing the recommender's predictions with users' answers shows that users do take the additional factors into account, when this information is available to them. Consequently, taking into account additional factors improves prediction accuracy. 
Our experiments show that the best results are obtained by combining content-based factors and additional factors in a proportion of about $10:4$, and differentiating the relevance of each additional factor
We also determined that users are not good judges of how additional factors impact their choices. Letting them decide the coefficients for the additional factors actually provides worse performance than having such coefficients learned for the whole population by the training algorithm.


The rest of the paper is structured as follows. We begin by positioning our work within pertinent research in Section~\ref{sec:related}.
In Section~\ref{sec:approach} we review the concepts and methodology we use, giving details of how we model events and how we represent users, and we introduce the linear scoring function, together with the different scoring factors, that we exploited in our study.
Section~\ref{sec:paramSetting} describes how we trained and tested the different scoring functions. Test results are compared and discussed in relation to research questions RQ1-RQ5 in Section~\ref{sec:discussion}. Section~\ref{sec:conclusions} concludes the paper.

\section{Related work }
\label{sec:related}
In this paper we investigate the relevance of social and contextual factors in predicting the interests of users in events and activities, with the ultimate goal of evaluating the usefulness and effectiveness of these features for automated recommendation. 

In general the idea of considering contextual and social factors in recommender systems is not new. However, this is seldom supported by studies determining their relevance w.r.t.\ the recommendation process. The contribution of this paper goes in this direction.

In this section, we first introduce  relevant work dealing with feature analysis, evaluation and selection for recommendation purposes. Then, we  briefly survey existing event recommenders, comparing the additional factors they include in their recommendation process. Finally, we present related areas of research: context-aware recommenders and multi-criteria recommenders.  

\subsection{Studies about additional factors in recommendation process}

Among a few research articles on recommender systems presenting a thorough user study, \cite{Baltrunas:11} propose the approach the most relevant to our work. The authors tackle the problem of recommending points of interest in the context of a digital travel guide and they are interested in finding out whether certain contextual factors (e.g.\ rainy weather, presence of children, etc.) have a positive, neutral or negative impact on users. They adopt an approach similar to ours: they ask users to rate some points of interest several times, the first time ignoring the contextual factors, the other times being aware of some of them. Differences in the ratings are used to estimate the impact of contextual factors, and the results are implemented in a mobile recommender system, which is then evaluated again against users' opinions. This work, among other things, shows the practical advantage of learning the effect of contextual factors by asking users to rate items in an imagined situation with different contexts. This approach has in fact the great advantage of reducing the cost of the context-dependent rating acquisition phase. Although the approach used in this work is quite similar to ours, there are two significant differences. First, the set of analyzed factors is substantially different: the only element in common is the distance between the user and the recommended object. This is partly due to the fact that \cite{Baltrunas:11} essentially recommend places, rather than activities, and that in this work they do not take into account the Social Web dimension. The second difference is that \cite{Baltrunas:11} evaluate the impact of contextual factors individually for each user, so that the newly acquired information becomes a part of the user model. Since their work is based on a recommender exploiting collaborative filtering techniques, this was the only possibility to incorporate contextual information. In our approach, we rather chose to evaluate the average impact of additional factors on a population. The resulting information was then directly incorporated in the recommendation algorithm, and tested on a different population.

\cite{Arazy:2010} propose a user study which focuses exclusively on social factors, and in particular on the type of social relationships existing among users. Starting from the fact, previously researched by others, that incorporating social information is beneficial only in a very limited set of cases, they performed a behavioral study of social factors and discovered that some types of social relationships (e.g.\ competence, benevolence) indeed affect user's decisions more than others. They then designed, developed, and tested a recommender system in order to verify how these factors impact the recommendation accuracy. Although this work deals with a different type of social factors than ours, their approach to the investigation supports our own, in that it follows the same outline: the recommendation process is tested under different hypotheses concerning the additional factors, and the improvement (or worsening) in the recommendation accuracy is used as a measure of the impact of the different factors.  

\cite{Quercia:2010} also analyze the impact of spatial and social factors (namely, distance and popularity) in the recommendation of events. Similarly to our approach, they compare six different scoring formulas that take into account these factors in different ways, and test them against user preferences to see which performs better. This allows them to conclude that recommendation based on event popularity {\em among those living in the event area} provides the best accuracy. Although this work may appear to be very similar to ours, there is a very significant difference: the authors base their study on real data concerning event attendance and users' localization, rather than on a survey as we did. This, in our opinion, rather than providing more realistic results, creates a sort of circularity in the argumentation: popularity is based on event attendance, but this information is available only {\em after} the event has taken place, when recommending it is useless. It could be argued that the authors are measuring the likelihood that a user attended an event based on how many other users in her neighborhood did. 

\cite{Odic13} try to decide which contextual information is relevant in the realm of context-aware recommender systems, since the relevant contextual information improves the recommendation quality, whereas the irrelevant contextual information can impact negatively the recommendation process. This work concerns the recommendation of movies, therefore the context is characterized in a different way than we propose in this paper.

 In general, to the best of our knowledge, there is no study concerning the social and contextual factors we discuss in the present paper. However, the above mentioned works show that the approach we followed is a viable one to the end of feature analysis and selection.

\subsection{Event recommender systems}
\label{sec:recSys}
In order to better contextualize our work and to motivate its relevance w.r.r.\ the state of the art in the recommendation of events, we briefly survey existing event recommenders, comparing the additional factors they include in their recommendation process. 

Recommender systems are usually divided into two broad categories (with possible hybrid solutions): content-based recommenders~\cite{Pazzani:2007}, which base their suggestions on the match among items features and the users' interests in a user model, and collaborative-filtering recommenders~\cite{Schafer:07}, which base their suggestions on what liked to similar users.
 
Many recommender systems, regardless of the specific recommendation algorithm they implement (content-based (\cite{Pazzani:2007}), collaborative filtering (\cite{Schafer:07}) or hybrid (\cite{Burke:07})), use in their recommendation process some of the factors we analyzed. However, to the best of our knowledge, none of them (i) uses these factors in systematic manner, i.e.\ after an extensive analysis of their effect on the recommendation process, (ii) considers all these factors together as we did, (iii) takes into account the influence of these factors on each other and analyzes their mutual interaction.

Table~\ref{related} provides a brief list of existing recommender systems for events with the indication of which context (including both space and time) and social factors they consider in their recommendation process. The first row corresponds to our study. 

\begin{table*}[t]
\centering
{\footnotesize
\begin{tabular}{lll}
&  \textbf{context factors} & \textbf{social factors} \\
\toprule
Our approach   & reachability   &  ratings of all users, friends behavior   \\
\midrule
PITTCULT~\cite{Lee:08}   &    &  trust   \\
\midrule
\cite{Zheng:10}   &  location, time  & user ratings of all users  \\ 
\midrule
CUPID \cite{Pessemier12} &  location  &  user participation \& ratings of all users \\  
\midrule
\cite{Minkov:10}  &  location  &  past user ratings  \\ 
\midrule
iCITY \cite{Carmagnola:2008} &  location    &  \\
\midrule
\cite{Cornelis:07}  &    &  ratings of similar users  \\ 
\midrule
\cite{Kayaalp:10} &   location   &   \\
\midrule
\cite{Waga:11} & location, time  &  ratings of all users  \\
\bottomrule
\end{tabular}
}

\caption{Factors used by the analyzed recommender systems}
\label{related}
\end{table*}

PITTCULT\footnote{\textit{http://pittcult.sis.pitt.edu/}} (\cite{Lee:08}), a recommender for cultural events, exploits social factors in terms of trust relations among users. PITTCULT applies standard collaborative filtering techniques, but rather than finding similar profiles among all users, it exploits trust mechanisms, where people can explicitly declare their trust toward other users. 

\cite{Zheng:10} focus on context factors in collaborative filtering setting, such as location and time zone information for users as well as for hotels, trip type and user's ratings and use them for travel recommendation. They identify important contexts for different algorithm components and propose to relax these contextual constraints for each algorithm component in order to improve the recommendation results.

CUPID (\cite{Pessemier12}), an event recommendation platform for the Flemish cultural scene, exploits an advanced collaborative-filtering algorithm extended with content-based filters to the pool of possible items to be recommended, in order to exclude undesired items, in accordance with the user profile. They use spatial factors to filter the results, excluding events that are too far, whereas the social factors (event popularity and user ratings) are used in the collaborative filtering phase to find similar users. 

\cite{Minkov:10} propose a collaborative ranking of future events (scientific talks) based on a hybrid recommender, which uses users' similarity based on their past feedback on the events (collaborative-filtering), the content of the event and location information (context information), but only for events within limited distances.

iCITY (\cite{Carmagnola:2008}) is an example of a content-based recommender providing information about cultural events in the city of Torino, Italy. Users can choose whether they prefer to have items recommended on the basis of their interests or on the basis of their location, however these two factors are not combined, and social aspects are not considered. 

\cite{Cornelis:07} propose a hybrid approach where a content-based system is extended with a collaborative dimension in the context of event recommendation, by positioning the users within a network of related individuals and allowing them to explore new items which those related users have appreciated in the past. 
In this work, the authors consider social features in the form of opinion of similar users on items of a domain and content features in the form of descriptions of items. They do not consider contextual features (such as distance).

In the work of \cite{Kayaalp:10}, event similarity is calculated combining content-based and collaborative filtering techniques. The authors also use the users' locations, which are calculated from their IPs in order to take into account users' reach ranges (similar to our willingness to move). This system is implemented as a Facebook application, where users can retrieve the events with all the corresponding information, including other users' ratings and the list of users that the same event was recommended to. In this case, such social factors are provided to the user but not considered in the recommendation process. 

In the work of \cite{Waga:11}, a context-aware recommender\footnote{In context-aware recommenders (\cite{Adomavicius:2011}), the activities or events are ranked by proximity, assuming that the user can be geo-localized, or by time of day depending on what is available at what time.} of location-based activities is described. Four relevant factors are considered in recommendation: content (free-form text description of photos), time (when the photo was taken), location (GPS coordinates of the user and of the photo content) and social network (i.e.\ other usersÕ ratings). They also rely on users' profiles by monitoring users' behavior and visited locations. Their scoring function is used to rank the services offered to users and it takes into account search history, location and explicit users' ratings. Regarding the location, they recommend only the resources in the immediate user proximity, while we also take into account the possibility for users to move. \\

\subsection{Pertinent areas of research}
A related area is indeed the one of \emph{context-aware recommenders}, since we focus, beside social, to contextual factors. 

In addition to \cite{Quercia:2010}, \cite{Odic13} and \cite{Waga:11}, we briefly present a few additional works in the field of \emph{context-aware recommenders}, even though they are not recommending events but related items like POI's or restaurants. 
\cite{Biancalana13}  present \emph{Polar}, a mobile social recommender which provides personalized recommendations of POI's in the vicinity of the user, taking into account the user preferences, current context (user location and activity, means of transportation, weather), and the information from her social network, user reviews and local search sites. The suggestions are filtered and ranked in order to meet user's needs.
\cite{Savage:12} designed mobile location-based recommender which infers user preferences from her social network profile by considering the contextual information related to the places the user visited in the past. It also uses mobile phone sensors for determining the user location and means of transportation (which indicates user's willingness to move).  
In order to enrich the recommendation process of restaurants, \cite{Lee06} present a recommender system which uses explicitly provided user preferences and three kinds of contexts: location (they consider the distance of the recommended item from the user's current position), personal context (information provided by the user model) and environment context. But the social aspect is not considered.\\

Another related field of research is definitely the area of \emph{multi-criteria recommender systems} (\cite{Adomavicius11,Manouselis07}), where users can provide their ratings not only on an item as a whole, but also on various attributes of the item. This brings improvements to the recommendation process since the suitability of an item for a certain user can be strongly influenced by these additional attributes of the item and user's complex preferences in that respect. 
In our study, when we asked the users to indicate how much the additional factors influence their intention to participate at an event, this could be seen as indirectly asking them to give ratings to these particular dimensions of the events. 
In multi-criteria recommender systems, very often aggregation function approach is used to predict the overall rating for an item, based on multi-criteria ratings. The aggregation function can be designed using expert domain knowledge or machine learning techniques (such as linear and non-linear regression or artificial neural networks). Our scoring function can be seen as linear combination of content and additional factors and our coefficients are estimated based on user ratings.  

\section{Methodology and concepts}
\label{sec:approach}

In order to find answers to the research questions raised in Section~\ref{sec:introduction}, we introduce a {\em scoring function} $\sigma(u,o)$ whose purpose is to predict the interest of a user $u$ in event $o$. Our scoring function is a linear function of the {\em scoring factors} that represent the different features and attributes that may take part in such a prediction. Its general form is thus given by the Equation~\ref{eq:linfun}:

\begin{equation}\label{eq:linfun}
\sigma(u,o) = w_0 + \sum_{j=1}^{m}w_j \cdot f_j(u,o)
\end{equation}

\noindent where $m$ is the number of scoring factors considered, $f_{1}(u,o), \ldots, f_{m}(u,o)$ are the scoring factors, $w_{1}, \ldots, w_{m}$ are the weights assigned to these scoring factors and $w_{0}$ is the error variable. 
Our goal is to learn the coefficients $w_0,w_1,\ldots, w_m$ for different combination of factors, by means of linear regression (\cite{witten2005datamining}). Given a set of $n$ training samples $(\overline{f_1}^{\textit{(i)}},\ldots,\overline{f_m}^{\textit{(i)}},\overline{\sigma}^{\textit{(i)}}), i=1, \ldots, n$, linear regression learns coefficients $w_0,\ldots, w_m$ in order to minimize the root mean square error (RMSE) given by~\ref{eq:RMSE}:

\begin{equation}\label{eq:RMSE}
\sqrt{\frac{\sum_{i=1}^{n}\left(\overline{\sigma}^{\textit{(i)}} - (w_{0}+\sum_{j=1}^{m}w_j\cdot\overline{f_j}^{\textit{(i)}}) \right) ^2}{n}}.
\end{equation}

The key steps of our approach, which will be detailed in the rest of this section and in the next one, are thus:
\begin{itemize}
\item Representation of events and users, detailing the information needed to compute the scoring factors (Sections \ref{sec:obj-model} and \ref{sec:user-model}).
\item Representation of each scoring factor as a formula computing a score from event and user features (Section~\ref{sec:scoring-factors}).
\item Training of different scoring functions, each corresponding to a certain combination of factors (Section~\ref{sec:training}). 
\item Testing of the scoring functions and comparison of the coefficients associated to the different factors, as a measure of their relevance for the final score (Section \ref{sec:testing}). 
\end{itemize}

Our approach can be seen as an instance of Multi-Attribute Utility Theory (MAUT). Our \emph{task} is to decide which additional factors to consider for good event recommendation and with which weights. We consider a number of \emph{alternatives} given by our initial assumptions (Section~\ref{sec:paramSetting}) where scoring factors (Themes, Type, Rating, Reachability, Friends' participation) represent MAUT \emph{dimensions}. Weights for dimensions are calculated using linear regression.

\subsection{Events}
\label{sec:obj-model}

We denote by $O$ the set of all the objects available for recommendation and by $U$ the set of all users. The properties we model for each event $o$ are the following:

\begin{description}
\item[Content properties] which are meant to capture the nature of the activity. 
In this work we characterize with labels the {\em types} and the {\em themes} of the activity. If, for example, we consider the domain of food-related events, \textsf{dinner} and \textsf{tasting} are types, while \textsf{fish} and \textsf{cheese} are themes. We associate with each event $o$ a non-empty set $\textit{THM}_o$ of theme labels and a non-empty set $\textit{TYP}_o$ of type labels. We denote by $\overline{\textit{THM}}$ and $\overline{\textit{TYP}}$ the set of all available labels for themes and types, respectively, and we assume these two sets to be disjoint.

\item[Spatial properties] which define the area $\textit{loc}_o$ where the activity takes place, described as a circle $(\hat{c}_o, \hat{r}_o)$ with center $\hat{c}_o = (\hat{x}_o,\hat{y}_o)$ and radius $\hat{r}_o$. 

\item[Social properties] which capture the interest expressed by other users within the same community in an event  $o$. We consider two social properties: ratings and participation. \\
\emph{Ratings} can be expressed as a partial function $\textit{rating}:U \times O \rightarrow I_{\textit{rat}}$, 
where $I_{\textit{rat}}=\left[ \textit{lb}_{\textit{rat}},\textit{ub}_{\textit{rat}} \right]$ is a closed interval over $\mathbb{R}$, $\textit{lb}_{\textit{rat}}$ and $\textit{ub}_{\textit{rat}}$ being its lower and upper endpoints. From now on we will always use $\textit{lb}$ and $\textit{ub}$ with the appropriate subscripts to denote the end points of a closed interval over $\mathbb{R}$.
The value $\textit{rating}(u,o)$ expresses how much user $u$ finds $o$ potentially interesting, without having a direct knowledge of $o$, since we assume $o$ has not yet taken place.
We also define the set $\textit{Raters}(o)$ of those users who have provided a rating for $o$. \\
\emph{Participation} corresponds to subset $\textit{Part}(o) \subseteq U$ of users that have confirmed their participation in $o$. 
\end{description}

\textbf{Example} 
Let us consider \emph{Salone Internazionale del Gusto},\footnote{Salone Internazionale del Gusto is a large, 5-day fair on the theme of sustainable and quality food taking place every two years in Torino, Italy. It organizes and promotes events such as dinners, tastings and debates, taking place on the entire region of Piemonte. \textit{http://salonedelgustoterramadre.slowfood.com/}} which took place at \emph{Lingotto Fiere exhibition center} in Torino, Italy, in October 2012. 
During this exhibition, a special dinner \emph{Belgian beers and sea food} was organized on \emph{Sunday, 28th of October, 2012}.  
The dinner took place at \emph{Eporidium} restaurant in Ivrea (Italy), 77 km north from \emph{Lingotto Fiere}. 

The event has been rated (on a scale 0-10) by 4 people, Mike (vote ``3''), Mark (vote ``6''), Mary (vote ``7'') and Megan (vote ``8'') and 3 people (John, Joseph and Jane) have confirmed their participation.

Let us denote the above described dinner by $o'$. Its properties can be formalized as follows:

\begin{itemize}[noitemsep]
\item \emph{Content properties}: 

$\textit{THM}_{o'} = \{fish, beer\}$

$\textit{TYP}_{o'} = \{dinner\};$

\item \emph{Spatial properties}: they are expressed as the circle $\textit{loc}_{o'}$, which covers the area of the \emph{Eporidium} restaurant where the activity took place, and which has center $\hat{c}_{o'} = (\hat{x}_{o'},\hat{y}_{o'})$ and radius $\hat{r}_{o'} = 0.1\, km$; $\hat{x}_{o'},\hat{y}_{o'}$ refer to the adopted coordinate system, e.g.\ GPS;

\item \emph{Social properties}: the set containing 4 users who rated $o'$ is $\textit{Raters}(o') = \{Mike, Mark, Mary, Megan\}$ and their rates are summarized in Table~\ref{table:socprop}:
\begin{table}[h]
\begin{center}
{\footnotesize
\begin{tabular}[t]{lc}
\toprule
 & rating \\
\otoprule  
 Mike & 3  \\
\midrule 
 Mark & 6  \\
\midrule 
 Mary & 7  \\
\midrule 
 Megan & 8  \\
\bottomrule
\end{tabular}}
\end{center}
\caption{Social properties of $o'$}\label{table:socprop}
\end{table}

Moreover, $\textit{Part}(o') = \{John, Joseph, Jane\}$ is the set of people who have subscribed to the event.

\end{itemize}

\subsection{User profile}
\label{sec:user-model}
The user profile is a knowledge structure capturing user features 
relevant in the recommendation process (such as demographics features, or user interests and preferences). In this section, we describe the features we need in order to compute our scoring factors. A recommender system considering these factors should similarly require these features in its user profile. 

Notice that we are not concerned here with how this information is determined or gathered: it may be partly explicitly provided by the users themselves, and partly inferred via some learning strategy taking into account interaction history. As each recommender system can implement its own strategy for information gathering and updating, this discussion is out of the scope of this paper.

We therefore assume that, for each user $u$ in the set $U$ of all users, the user profile contains the following information: 

\begin{description}
\item[Spatial properties:] they include {\em user's position in space} and {\em willingness to move}. 
We represent the user's position similarly to event locations, as a circle $(\hat{c}_u, \hat{r}_u)$ with center $\hat{c}_u = (\hat{x}_u,\hat{y}_u)$ and radius $\hat{r}_u$. 
We adopt an area, rather than a point, because for users it is not always possible to define a point that describes where the ``user position'' indeed is, depending on the presence and precision of the geo-localization system (GPS, Wi-Fi, etc.). For example, the position can correspond to a generic ``home'' location, or be provided explicitly. Moreover, the area in which the user wants to obtain recommendations may vary, depending on the situation. For example, if the user is not geo-positioned, but we simply know she is in ``San Francisco'', than $\hat{c}_u$ and $\hat{r}_u$ are estimated to be the center and radius of San Francisco's metropolitan area. 

The willingness to move is expressed as a distance measure $\textit{mov}_u$ which tells us the maximum distance the user is willing to travel in order to participate in an activity. {Notice that this, also, can be inherent to the specific situation. In their everyday life people may not want to cover long distances for an evening out, while on a holiday trip they could feel more explorative.}

\item[User interests:] it describes the interest of a user for the types and themes of events, as defined above. For our purposes, it suffices to represent interest as a function $\textit{int}: U \times (\overline{\textit{THM}} \cup \overline{\textit{TYP}}) \rightarrow I_{\sigma}$, where $I_{\sigma}= \left[\textit{lb}_\sigma, \textit{ub}_\sigma\right]$ is a closed interval over $\mathbb{R}$, 

\item[Social network:] it represents the network of friendships among users. Although this information can be expressed at different levels of complexity, for the present work we are only concerned with the friendship relation, seen as a binary, reflexive, and possibly non-symmetrical relationship $F \subseteq U \times U$: if $(u,v) \in F$ then $u$ has declared $v$ to be her friend (and not necessarily vice-versa). Therefore, for a given user $u$, we can define the set of her friends $\textit{F}_u \subseteq U$ as $\textit{F}_u = \{ v \mid (u,v) \in F\}$.
\end{description}

\textbf{Example} Consider Susan (user $u'$), who is visiting \emph{Salone Internazionale del Gusto}. She is in the area of \emph{Lingotto Fiere exhibition center} in Torino during the five days of the fair, and is willing to travel for not more than 100 km to reach interesting events near Torino. The fair offers different kinds of activities (workshops, tastings, debates and dinners) about a number of subjects (fish, coffee, wine, beer, cheese and cold cuts). Susan loves dinners and tastings, but she is not interested at all in other activities. Moreover, she loves cheese, fish and cold cuts, and she likes coffee but she does not drink alcohol. Finally, three of Susan's friends are going to visit \emph{Salone Internazionale del Gusto}, namely John, Joseph and Kate. Formally, Susan can be described as follows:
\begin{itemize}[noitemsep]
\item \emph{Spatial properties}: the circle $(\hat{c}_{u'}, \hat{r}_{u'})$ with center $\hat{c}_{u'} = (\hat{x}_{u'},\hat{y}_{u'})$ and radius $\hat{r}_{u'} = 0.0\, km$ describing her position refers to a fixed point at the \emph{Lingotto Fiere exhibition center}. The coordinates refer to the same coordinate system adopted for the spatial properties of the events. Susan's willingness to move is $\textit{mov}_{u'} = 100\, km$. 

\item \emph{Profile}: Susan's profile, in terms of her interest w.r.t.\ the themes and types on a scale 0-10, is described in the Table~\ref{table:susprof}:

\begin{table}[h]
\begin{center}
\footnotesize{
\begin{tabular}[t]{cc}
\toprule
 Theme & Interest \\
\otoprule  
 fish & 10  \\
\midrule 
coffee & 7  \\
\midrule 
 wine & 0  \\
\midrule 
 beer & 0  \\
\midrule 
 cheese & 8  \\
\midrule 
 cold cuts & 9  \\
\bottomrule
\end{tabular}
\hspace{1cm}
\begin{tabular}[t]{cc}
\toprule
 Type & Interest \\
\otoprule
 workshop & 2  \\
\midrule 
 tasting & 10  \\
\midrule 
 debate & 1  \\
\midrule 
 dinner & 9  \\
\bottomrule
\end{tabular}
}
\end{center}
\caption{Susan's profile}\label{table:susprof}
\end{table}

\item \emph{Social network}: 

$\textit{F}_{u'} = \{John, Joseph, Kate\}.$
\end{itemize}

\subsection{Scoring factors}
\label{sec:scoring-factors}
The main task of the scoring function is to assign a score to a pair $(u,o)$, where $u$ is a user and $o$ is an object to recommend, which expresses an estimate of how much the user $u$ would be interested in participating in the event $o$. 
The score can be thus represented as a function $\sigma:U \times O \rightarrow I_{\sigma}$, where  $I_{\sigma} = \left[\textit{lb}_\sigma, \textit{ub}_\sigma\right]$ is a closed interval over $\mathbb{R}$. 

Each scoring factor, in its more general form, is a function $f: U \times O \rightarrow \mathbb{R}$. In order to apply linear regression we do not need to put any constraint on the size of each factor. However, if we want to compare the coefficients learned through linear regression to infer the relevance of the different factors, they need to range over similar intervals. If, for example, a factor $f_1$ ranges over $[0-100]$, while factor $f_2$ ranges over $[0-5]$, it is difficult to conclude that a lower coefficient for $f_1$ means that $f_1$ is less relevant.

For this reason we decided to scale every scoring factor that ranges over a closed interval so that it is mapped to the interval $I_{\sigma}$. This mapping can be achieved by applying a scaling function \textit{map} which maps the codomain of function \textit{f} into the codomain $I_\sigma$ of the scoring function. A scaling function $\textit{map}_{A \rightarrow B}$ from a real interval 
$A=[\textit{lb}_A,\textit{ub}_A]$ to a real interval $B=[\textit{lb}_B,\textit{ub}_B]$
can in general be expressed with the Equation~\ref{eq:map}:

\begin{equation}\label{eq:map} 
\textit{map}_{A\rightarrow B} (x)= (x-\textit{lb}_A)\frac{\textit{ub}_B - \textit{lb}_B}{\textit{ub}_A - \textit{lb}_A} + \textit{lb}_B.
\end{equation}    

This poses a problem for scoring factors that do not range over a closed interval; in our case this is true only for factor {\em frn}, which denotes friends' participation. We will discuss how we deal with this issue when we describe the corresponding function.

The scoring factors we consider can be expressed as follows:

\begin{description}
\item[Thematic interest:] $\textit{thi}:U\times O \rightarrow I_{\sigma}$ is defined as the average interest expressed by user $u$ (stored in the user profile) for the themes of event $o$ and can be expressed with Formula~\ref{eq:thi}:
\begin{equation}\label{eq:thi}
\textit{thi}(u,o) = \textit{map}_{\textit{int}\rightarrow\sigma}\left(\frac{\sum_{t\in\textit{THM}_o}\textit{int}(u,t)}{|\textit{THM}_o|}\right).
\end{equation}

\item[Type interest:] $\textit{tyi}:U\times O \rightarrow I_{\sigma}$ is analogous to thematic interest, but takes into consideration the types of event rather than its themes and can be expressed with Formula~\ref{eq:tyi}: 
\begin{equation}\label{eq:tyi}
\textit{tyi}(u,o) = \textit{map}_{\textit{int}\rightarrow\sigma}\left(\frac{\sum_{t\in\textit{TYP}_o}\textit{int}(u,t)}{|\textit{TYP}_o|}\right).
\end{equation}

\item[Average rating:] $\textit{rat}: O \rightarrow I_{\sigma}$ computes the average rating of all the users in the community for a given event. This scoring factor depends only on the event and not on the user $u$ whom we are computing the score for and is given by Formula~\ref{eq:rat}: 
\begin{equation}\label{eq:rat}
\textit{rat}(o) = \textit{map}_{\textit{rat}\rightarrow\sigma}\left(\frac{\sum_{\textit{v}\in\textit{Raters}(o)}\textit{rating}(v,o)}{|\textit{Raters}(o)|}\right).
\end{equation}

\item[Reachability:] $\textit{rch}:U\times O \rightarrow I_{\sigma}$ captures the willingness of a user $u$ to cover the existing distance between herself and the event. It is not a mere distance measure, but it gives a sort of user's judgment on how distant she considers the event and on how ready to move she is to reach it. In fact, it linearly decreases until the distance between the user's and event positions reaches the user's declared {\em willingness to move}. Figure~\ref{fig:rchFunc} shows $\textit{rch}$ as a function of the distance  $\textit{dist}(\hat{c}_u,\hat{c}_o)$ between the user's and the event area centers. Given the upper and lower bounds of our scores, $\textit{ub}_\sigma$ and $\textit{lb}_\sigma$, we want (i) $\textit{rch}(u,o)$ to be maximal ($\textit{rch}(u,o) = \textit{ub}_\sigma$) when user's area center $\hat{c}_u$ and event area center $\hat{c}_o$ coincide, and (ii) $\textit{rch}(u,o)$ to be minimal ($\textit{rch}(u,o) = \textit{lb}_\sigma$) when the distance between user's area and event's area is equal or greater than the user's {\em willingness to move} ($\textit{dist}(\hat{c}_u,\hat{c}_o) \geq\hat{r}_u + \hat{r}_o + \textit{mov}_u$, see Figure~\ref{fig:rch}).

This can be obtained by using the following piecewise linear Function~\ref{eq:rch}:
\hspace*{3mm}
\begin{equation}\label{eq:rch}
\textit{rch}(u,o) =  \max \left(\textit{lb}_\sigma, \frac{\textit{lb}_\sigma - \textit{ub}_\sigma}{\hat{r}_u + \hat{r}_o + \textit{mov}_u} \cdot \textit{dist}(\hat{c}_u,\hat{c}_o) + \textit{ub}_\sigma\right).
\end{equation}

\begin{figure}[ht]
\begin{center}
\includegraphics[width=\linewidth]{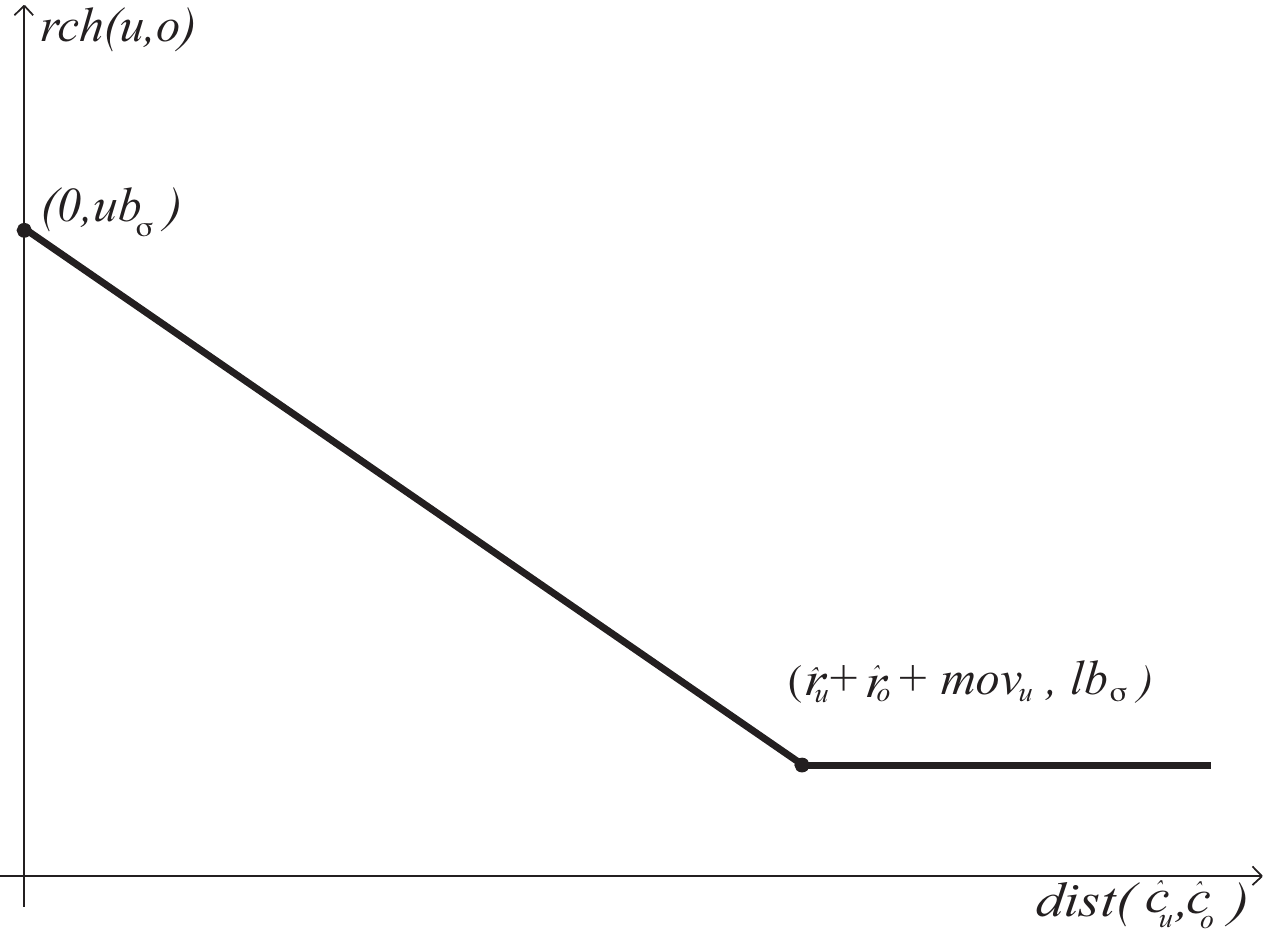}
\caption{How the reachability function decreases with distance}
\label{fig:rchFunc}
\end{center}
\end{figure}

\begin{figure}
\begin{center}
\includegraphics{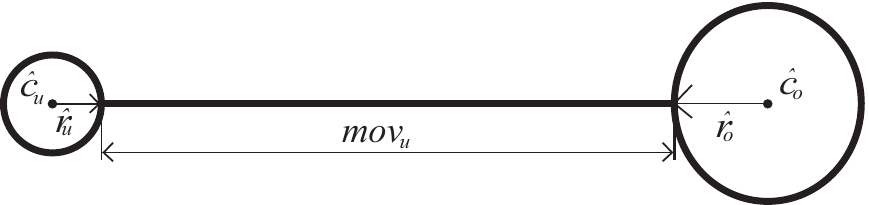}
\caption{Reachability is minimal when the distance between user's area and event area is equal or greater than the user's willingness to move $\textit{mov}_u$}
\label{fig:rch}
\end{center}
\end{figure}

\item[Friends' participation:] $\textit{frn}:U\times O \rightarrow I_{\sigma}$ gives a measure of how many of user's friends participate in the event. 
To account for a significant increase in the scoring factor for small numbers, and a slow increase for larger ones, we decided to define the scoring factor as a logarithmic function of the actual number of participating friends (plus 1, to handle the case where the number of friends is 0).

As we mentioned above, this poses a problem, since the codomain of a logarithmic function of this type is $[0,\infty)$ and it is not possible to map it to $I_{\sigma}$. In order to scale this factor to make it comparable to the others, we decided to define friends' participation with the following Formula~\ref{eq:frn}:
\begin{equation}\label{eq:frn}
\textit{frn}(u,o) = \textit{lb}_\sigma + \frac{\textit{ub}_\sigma - \textit{lb}_\sigma}{\ln(K+1)} \ln(|PartF_{u}(o)| + 1),
\end{equation}

where $PartF_{u}(o) = \textit{Part}(o) \cap F_u$ and $K$ is the number of friends for which we want to obtain that $\textit{frn}(u,o) = \textit{ub}_\sigma $. By defining the function in this way, for numbers of friends larger than $K$ the function grows very slowly, so that there are very few values significantly larger than $\textit{ub}_\sigma$.
 
In order to choose a value for $K$, we conducted an additional survey (independent from those used to collect training and test samples) to understand the users' attitude toward the number of friends participating in the events. We wanted to discover whether ``the more the better'' (so to speak), or whether from a certain number on people actually felt it did not make any difference. We asked people to associate specific numbers to the following expressions: ``I am going with a couple of friends'', ``with some friends'', ``with a group of friends'', ``with many friends''. By analyzing 91 answers, we concluded that people differentiate quite clearly among these expressions. On  average they consider more than 8 friends just ``many'', while 4-8 friends is ``a group'', 3-5 friends is ``a few'' and 2-3 friends is ``a  couple''. Therefore we set $K=8$.

\end{description}

Finally, the \textbf{global scoring function} can be expressed as the weighted sum of the scoring factors described above and is given by the following Formula~\ref{eq:scorefun}:

\begin{multline}\label{eq:scorefun}
 \sigma(u,o) = \omega_{\textit{thi}} \cdot \textit{thi}(u,o)  + \omega_{\textit{tyi}} \cdot \textit{tyi}(u,o) + \omega_{\textit{rch}} \cdot \textit{rch}(u,o) \\ 
 + \omega_{\textit{rat}} \cdot \textit{rat}(o) + \omega_{\textit{frn}} \cdot \textit{frn}(u,o) + \omega_{0}.
\end{multline}

\textbf{Example}

As an example, consider Susan's interest in the dinner \emph{Belgian beers and sea food}. Recall that $\textit{THM}_{o'} = \{fish, beer\}$ and $\textit{TYP}_{o'} = \{dinner\}.$ Let us moreover assume that our scoring function ranges over $[0,10]$, thus $\textit{lb}_\sigma = 0$ and $\textit{ub}_\sigma = 10$.

The thematic interest is:
$$
\textit{thi}(u',o') = \frac{10+0}{2} = 5,
$$  
the type interest is:
$$
\textit{tyi}(u',o') = 9.
$$  
Also recall that $\textit{Raters}(o') = \{Mike, Mark, Mary,$ $Megan\}$ and their respective rates are ``3'', ``6'', ``7'', ``8''. Then
the average rating is
$$
\textit{rat}(o') = \frac{3+6+7+8}{4} = 6.
$$

Referring again to Susan ($u'$) and the dinner \emph{Belgian beers and sea food} ($o'$), the reachability $\textit{rch}(u',o')$ of $o'$ for $u'$ is 
$$\textit{rch}(u',o') = \max \left(0,\frac{0-10}{0+0.1+100} \cdot 77 + 10\right) = 2.31$$
\noindent since $\hat{r}_{u'} = 0$, $\hat{r}_{o'} = 0.1$, $mov_{u'} = 100\, km$ and $dist(\hat{c}_{u'},\hat{c}_{o'})=77\, km$.
To conclude our example, two of Susan's friends participate in the dinner \emph{Belgian beers and sea food}: $Part(o') = \{John, Joseph, Jane\}$ and $F_{u'} = \{John, Joseph, Kate\}$. Therefore the score for friends' participation is:
\begin{align*}
\textit{frn}(u',o') 
&= 0 + \frac{10-0}{\ln(9)} \cdot \ln(2+1) = 4.5512 \cdot ln(3) = 5.
\end{align*}

\section{Data analysis}
\label{sec:paramSetting}

In order to investigate the issues stated in Section \ref{sec:introduction}, we learned the coefficients for the scoring function \eqref{eq:scorefun} using linear regression, under different initial assumptions, actually obtaining a set of different scoring functions. By comparing {\em (i)} the capability of these different functions to predict the user's interest in an event, and {\em (ii)} the coefficients learned for each scoring factor, we then discuss possible answers to our research questions RQ1-RQ5.

We learned the scoring function under four different sets of initial assumptions, thus obtaining several variants with different coefficients:

\newcommand{\ic}{\ensuremath{\mathcal{IA}_{0}}}
\newcommand{\ica}{\ensuremath{\mathcal{IA}_{X}}}
\newcommand{\icu}{\ensuremath{\mathcal{IA}_{Xu}}}
\newcommand{\icd}{\ensuremath{\mathcal{IA}_{Xd}}}

\begin{description}
\item[\ic:] only the factors concerning theme and type interest (\textit{thi} and \textit{tyi}) are considered;
\item[\ica:] all five factors are considered;
\item[\icu:] all five factors are considered, but the relative coefficients of the three additional factors are assumed to be given by the preferences explicitly expressed by the user;
\item[\icd:] all five factors are considered, however we learn nine different functions, each applicable only to users whose theme and/or type interest fall into a certain range. We call this the {\em dynamic} approach. 
\end{description}


\subsection{Data collection}
\label{sec:datacoll}

In order to learn the scoring functions using linear regression, and subsequently test the results, we collected data from potential users. We designed two online surveys where people could express their interest in participating in certain events, all related to wine and food.

Data from the first survey (in the following, $S_{\textit{ten}}$) was used for training and validation, by performing a 15-fold random split into a training set with $2/3$ of the samples and a validation set with the remaining $1/3$. Data from the second survey (in the following, $S_{\textit{test}}$) was used as test set. 

In both surveys participants were asked to imagine they would spend a few days in Torino, Italy, on the occasion of the upcoming \emph{Salone Internazionale del Gusto}, and that they were lodged close to the main site where the \emph{Salone} takes place.
This allowed us to set the same {\em user position} for all users (namely $\hat{c}_u = (0,0)$ and $\hat{r}_u = 0.0\textit{km}$), thus considering a relative system of coordinates centered in the location of the fair. 

In each survey we proposed a set of 15 events among typical \emph{Salone} activities\footnote{In order to have enough events for each theme/type we actually selected activities from many different editions of the fair.}. Their relevant features are shown in Table~\ref{table:stobs}. Events $o_i$ were proposed in the first survey, while events $o'_i$ were proposed in the second survey. Each event was characterized by:
\begin{itemize}
\item $1-3$ themes and its type;
\item the distance between the user center $\hat{c}_u$ and the event center $\hat{c}_o$, and a fixed event radius $\hat{r}_o = 0.1$;
\item a made up ``number of friends'' that said they intended to participate in the event;
\item a made up average rating.
\end{itemize}
Notice that most of the activities took place in the main site of the \emph{Salone}; some of them were located in close vicinity (in the city of Torino), and a few others were further away, in the countryside. This is typical for this type of event.
\begin{table*}[hbt]
\begin{center}
\begin{tabular}{p{0,9cm}p{4,9cm}p{3cm}p{1,2cm}p{1,2cm}p{1,2cm}}
\toprule
Event & Themes& Type    & Distance [km] & Friends Nr. & Rating \\ \toprule
$o_1$ & fish      & workshop & 0      & 0   & 8  \\ 
$o_2$ & coffee    & workshop & 0      & 3   & 3  \\ 
$o_3$ & wine      & workshop & 36     & 1   & 8  \\ 
$o_4$ & beer, cheese      & workshop & 4      & 0   & 9  \\ 
$o_5$ & beer      & tasting  & 0      & 1   & 7  \\ 
$o_6$ & cheese, wine      & tasting  & 6      & 1   & 6  \\ 
$o_7$ & cold cuts & tasting  & 0      & 0   & 6  \\ 
$o_8$ & beer      & debate   & 6      & 2   & 8  \\ 
$o_9$ & wine      & debate   & 0      & 0   & 7  \\ 
$o_{10}$ & cold cuts & debate   & 0      & 5   & 5  \\ 
$o_{11}$ & coffee & debate & 22  & 4 & 8  \\
$o_{12}$ & fish, oil & dinner   & 6      & 0   & 9  \\ 
$o_{13}$ & fish, beer        & dinner   & 77     & 8   & 4  \\ 
$o_{14}$ & cheese    & dinner   & 3      & 3   & 6 \\
$o_{15}$ & cheese, cold cuts & dinner   & 0      & 4   & 4  \\ \bottomrule
$o'_1$  & beer      & meeting & 7      & 4   & 8  \\ 
$o'_2$   & meat, fruit and vegetables      & cooking course & 0      & 1   & 6  \\ 
$o'_3$  & chocolate, wine & dinner & 18  & 10 & 9 \\
$o'_4$   & oil, wine        & dinner   & 0     & 2   & 8  \\ 
$o'_5$    & cheese, oil & workshop   & 0      & 0   & 7  \\ 
$o'_6$    & beer, chocolate    & workshop & 0      & 12   & 6  \\ 
$o'_7$     & fruit and vegetables      & cooking course  & 0      & 6   & 4  \\ 
$o'_8$    & wine      & meeting  & 0      & 5   & 8  \\ 
$o'_9$   & fruit and vegetables, wine      & dinner   & 6      & 3   & 6  \\ 
$o'_{10}$  & fruit and vegetables & cooking course   & 22      & 1   & 9  \\ 
$o'_{11}$  & meat, wine      & workshop & 55     & 2   & 3  \\ 
$o'_{12}$  & meat, cheese      & cooking course & 0     & 6   & 10  \\ 
$o'_{13}$  & cheese, wine & meeting  & 0      & 0   & 7  \\ 
$o'_{14}$   & cheese, wine      & dinner   & 65      & 5   & 7  \\ 
$o'_{15}$  & oil, fruit and vegetables, wine  & workshop   & 0      & 2   & 5 \\ \bottomrule
\end{tabular}
\end{center}
\caption{Events in the experimental set up}\label{table:stobs}
\end{table*}

\begin{figure*}
\begin{center}
{\includegraphics[width=0.95\textwidth]{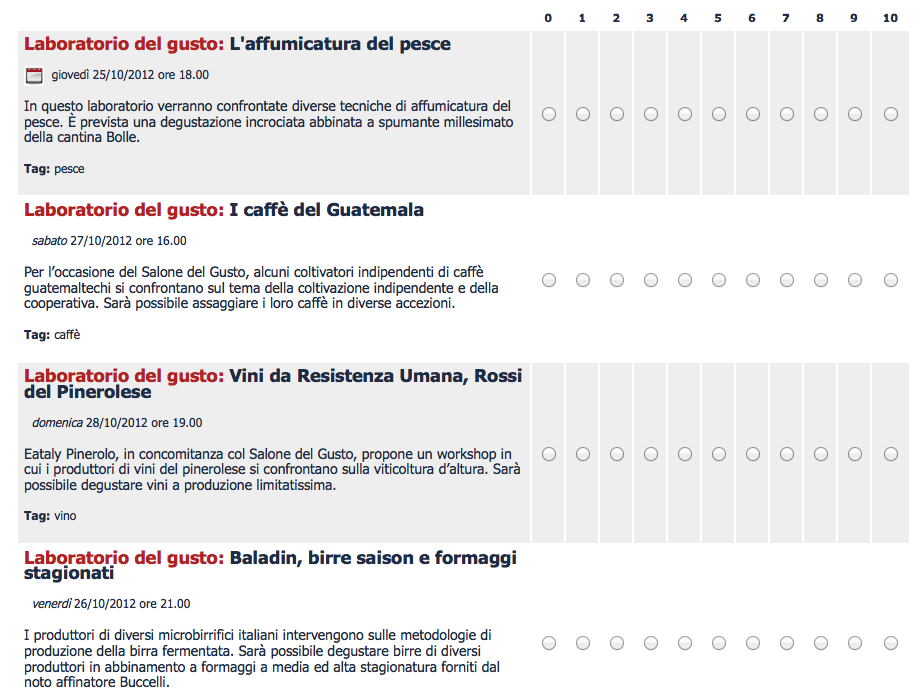}} \quad
\caption{A part of the page containing all the events for the {\em init} user scores.}
\label{fig:events}
\end{center}
\end{figure*}

The events were shown in a single scrollable web page of the questionnaire.  
Figure~\ref{fig:events} shows a part of the scrollable web page containing the 15 events of the first survey. 

\begin{figure}
\begin{center}
{\includegraphics[width=0.42\textwidth]{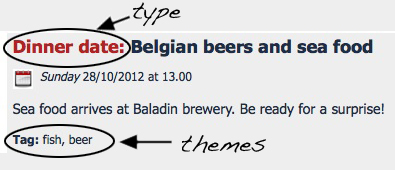}} 
\caption{Survey - an event with information limited to themes and type.}
\label{fig:surveyEvents1}
\end{center}
\end{figure}

\begin{figure}
\begin{center}
{\includegraphics[width=0.46\textwidth]{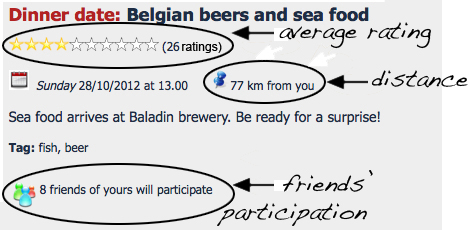}} \quad
\caption{Survey - an event with full information.}
\label{fig:surveyEvents2}
\end{center}
\end{figure}

In the first survey we asked each user to provide the following information :
\begin{itemize}
\item Age, social and tourist network usage (see Table \ref{tab:demo:grp}).
\item A score $0-10$ expressing interest in each event, \emph{knowing only its themes and type} (see Figure~\ref{fig:surveyEvents1}); these will be referred to as the {\em init} user scores, and they will be denoted by $\tau^0_{u}(o)$ for user $u$ and event $o$. 
\item A score $0-10$ expressing interest in each event, knowing its distance from the main fair site, its rating and the number of friends attending it, in addition to its themes and type (see Figure~\ref{fig:surveyEvents2}\footnote{Figures \ref{fig:surveyEvents1} and \ref{fig:surveyEvents2} were translated to English for readers' convenience, the original ones were presented to users in Italian.}); these will be referred to as the {\em fin} user scores, and $\tau_{u}(o)$ will denote such ``final'' scores for user $u$ and event $o$. 
\item A value $0-10$ expressing interest in the different themes and types.
\item Three values $0-10$ expressing a self-assessment on the relevance of distance, friend participation and rating when expressing interest for an event; we will denote these values as $W_{\textit{rch}}(u)$, $W_{\textit{frn}}(u)$ and $W_{\textit{rat}}(u)$ respectively.
\item The longest distance one was willing to cover to join a {\em Salone}-related event.
\end{itemize}

In the second survey, we did not ask the users for their scores expressing interest in events knowing only their themes and types. Although the two surveys are similar, it is important to point out that the second one was conceived after a few preliminary tests on the first data set. As we felt that on some specific issues we had already collected enough evidence, the second survey was shorter and less information was asked from the participants, in order not to burden them with a time-consuming task.

People who volunteered to answer the surveys were recruited on Facebook. The two surveys were published four months apart, and each remained open for about two weeks. The URLs of the surveys were made publicly available; people who chose to participate filled in the survey on their own (no supervision was required). Participants were thus recruited according to an availability sampling strategy.\footnote{Much research in social science is based on samples obtained through non-random selection, such as the availability sampling, i.e.\ a sampling of convenience, based on subjects available to the researcher, often used when the population source is not completely defined.} Recruiting users by means of a social networking site was intended to select potential target users for event recommendation within the same context. In fact, our sample has the demographical features distribution comparable with the Facebook users, according to some studies\footnote{http://www.socialbakers.com/facebook-overview-statistics/}. Table \ref{tab:demo:grp} reports demographic information for the three groups of users.

\begin{table}[h]
\begin{center}
\begin{tabular}{lccccc}
\toprule
\multicolumn{6}{c}{Age distribution}\\ \toprule
& & \textbf{18-29} & \textbf{30-39} & \textbf{40-49} & \textbf{50+} \\  \midrule
$S_{\textit{tr}}:$ & & $34\%$ & $47\%$ & $12\%$ & $7\%$\\ \bottomrule
$S_{\textit{test}}:$ & & $35\%$ & $47\%$ & $13\%$ & $5\%$\\ \bottomrule
\toprule
\multicolumn{6}{c}{Usage of social networking apps}\\ \toprule
& \textbf{never} & \textbf{rarely} & \textbf{sometimes} & \textbf{often} & \textbf{always} \\ \midrule
$S_{\textit{tr}}:$ & $0\%$ & $6\%$ & $18\%$ & $42\%$ & $34\%$ \\ \bottomrule
$S_{\textit{test}}:$ & $0\%$ & $6\%$ & $15\%$ & $43\%$ & $36\%$ \\ \bottomrule
\toprule
\multicolumn{6}{c}{Usage of tourist recommendation apps}\\ \toprule
& \textbf{never} & \textbf{rarely} & \textbf{sometimes} & \textbf{often} & \textbf{always} \\ \midrule
$S_{\textit{tr}}:$ & $18\%$ & $21\%$ & $33\%$ & $21\%$ & $7\%$ \\ \bottomrule
$S_{\textit{test}}:$ & $15\%$ & $23\%$ & $30\%$ & $21\%$ & $11\%$ \\ \bottomrule
\end{tabular}
\end{center}
\caption{Demographic information}
\label{tab:demo:grp}
\end{table}

The first survey, used for training and validation, was answered by 300 users, while the second survey, used for testing, was answered by 108 users. Therefore the training and validation sets share the same set of events, but concern completely different users, while the second test set presents a completely new set of events, but the set of users may include someone who was also included in the training or validation set, as the surveys were anonymous and we did not prevent users who participated in the first one to answer also the second one.
In order to learn the scoring functions, we needed to compute the scoring factors (\textit{thi}, \textit{tyi}, \textit{rch}, \textit{rat}, \textit{frn}) for each (user,event) pair. The learning and test process were then carried out on an updated version of the training and test sets, which included this information.

\subsection{Computing the scoring factors}\label{sec:factors}
 
Given the data in $S_{\textit{tr}}$ and $S_{\textit{test}}$, for each pair $(u,o)$ of a user $u$ and an event $o$ we computed the five scoring factors in the following way:
\begin{description}
\item[Thematic interest $\textit{\rmfamily thi}(u,o)$, type interest $\textit{\rmfamily tyi}(u,o)$:] we used the interest scores explicitly provided by users in our questionnaire. For events with two or three themes, we computed the average value using Formulas~\ref{eq:thi} and~\ref{eq:tyi}. As the scores provided by users are in the interval $[0,10]$, so are the resulting values for thematic and type interest. Therefore, they already belong to the final score interval $I_{\sigma}$. 
\item[Average rating $\textit{\rmfamily rat}(o)$:] we used the pre-determined value associated with each event. These are therefore simulated values. Also these values already belong to the interval $[0,10]$.
\item[Friends' participation $\textit{\rmfamily frn}(u,o)$:] we fed the pre-determined value associated with each event to the Formula~\ref{eq:frn} provided in Section \ref{sec:scoring-factors}, with $\textit{lb}_\sigma = 0$ and  $\textit{ub}_\sigma = 10$.
\item[Reachability $\textit{\rmfamily rch}(u,o)$:] we used the Formula~\ref{eq:rch} provided in Section~\ref{sec:scoring-factors}, with $\textit{lb}_\sigma = 0$ and  $\textit{ub}_\sigma = 10$, in order to again obtain a value in the same range as the other factors.
The distance between the centers of user's and event areas ($\textit{dist}(\hat{c}_u,\hat{c}_o)$) was in this case explicitly specified in the event description, while the user's willingness to move ($\textit{mov}_u$) was directly provided by the user as a part of the survey.
The radius of the event area ($\hat{r}_o = 0.1 \ km$) and the radius of the user's area ($\hat{r}_u = 0$) are predefined values in the models.
\end{description}

\subsection{Training}\label{sec:training}
Training was performed using the Weka Environment for Knowledge Analysis\footnote{http://www.cs.waikato.ac.nz/ml/weka/} (\cite{witten2005datamining}), selecting the standard Linear Regression algorithm for classification, 
with a ridge value of $1.0E-8$, since preliminary cross-testing did not suggest that regularization could improve the results. Linear Regression assumes that the samples are independent (differently from regression methods based on multilevel models); we consider this to be a reasonable assumption, given that each pair (user, event) occurs only once and therefore no item was rated twice by the same user.

Linear Regression in Weka allows to select the attributes that should be used for learning (in our case, these corresponded to the scoring factors) and which is the feature whose value should be predicted (in our case, this corresponded to the user score).

We recall that we learned several scoring functions, under different initial assumptions, which means using different sets of attributes. In the following, we describe the setup and outcome (in terms of the learned function) of each training process.

\begin{enumerate}
\item \textbf{\ic} - use only theme and type interest.
	\begin{description}
	\item[Assumption:] {\em the input attributes are {\em thi} (theme interest) and {\em tyi} (type interest). We learn to predict both {\em init}al and {\em fin}al user scores, obtaining two scoring functions $\sigma^{\textit{init}}_0$ and $\sigma^{\textit{fin}}_0$. In the first case the user has the same knowledge as the learning algorithm, namely only themes and type of the event are shown to her, whereas in the second one the user knows more than the learning algorithm, since the additional factors are also presented.}
	\item[Outcome:]\hfill\\
	$\sigma^{\textit{init}}_0 = 0.5911  \textit{thi} + 0.3338  \textit{tyi} - 0.2524$.\\
	$\sigma^{\textit{fin}}_0 =  0.597   \textit{thi} + 0.3235 \textit{tyi} - 0.8436$.
	\end{description}
	
\item \textbf{\ica} - use all five factors.
	\begin{description}
	\item[Assumption:] {\em we consider all five factors. In this case, as in the following ones, we learn to predict only {\em fin}al user scores, obtaining one scoring function $\sigma_{X}$ (we omit the superscript).}
	\item[Outcome:]  \hfill\\
	 $\sigma_{X} =   0.5698  \textit{thi} + 0.3286  \textit{tyi} + 0.0848  \textit{rat} + 0.1967  \textit{rch} + 0.07965  \textit{frn} - 3.0467$.
	\end{description}
	
\item \textbf{\icu} - use all five factors, with the coefficients for the additional factors based on user self-assessment.
	\begin{description}
	\item[Assumption:] {\em here the relevance of the three additional factors (rating, friends' participation and reachability) is computed according to the self-assessment given by users in the survey. Here we learn two scoring functions according to two slightly different interpretations of the user-provided coefficients. In the first case we considered them as absolute values on a $[0-10]$ scale; the combined scoring factor was then computed as $\textit{u-abs} = (W_{\textit{rat}}\cdot\textit{rat}+W_{\textit{rch}}\cdot\textit{rch} + W_{\textit{frn}}\cdot\textit{frn})/30$. In the second case we considered them as relative to each other; the combined scoring factor was computed as $\textit{u-rel} = (W_{\textit{rat}}\cdot\textit{rat}+W_{\textit{rch}}\cdot\textit{rch} + W_{\textit{frn}}\cdot\textit{frn})/(W_{\textit{rat}} + W_{\textit{rch}} + W_{\textit{frn}})$. We fed each of these new attributes to the training algorithm, together with theme and type interest.} 
	\item[Outcome:]  \hfill\\
	 $\sigma^{\textit{abs}}_{Xu} =   0.5835  \textit{thi} + 0.3199  \textit{tyi} + 0.2799  \textit{u-abs} - 1.6102$.\\
	 $\sigma^{\textit{rel}}_{Xu} =   0.5782  \textit{thi} + 0.3329  \textit{tyi} + 0.2331  \textit{u-rel} - 2.1925$.
	\end{description}

\item \textbf{\icd/thi} - use all five factors, dynamic approach depending on theme interest.
	\begin{description}
	\item[Assumption:] {\em the dynamic approach assumes that additional factors may have a different relevance depending on how much the user is interested in the theme and/or type of the event. We thus have three sub-hypotheses, depending on whether we consider the dependence with respect to theme interest, type interest or both. Here we are considering the first case.
		
	 In order to test this initial assumption, we split the samples in the training set into three groups, according to the value of factor {\em thi} (theme interest). These subgroups were created choosing two thresholds (namely, 6 and 8) that partition the sample set in three subsets of similar size (about 1000 samples each): a set of samples with $\textit{thi} \in [0,6)$, a set of samples with $\textit{thi} \in [6,8)$, and a set of samples with $\textit{thi} \in [8,10]$. 
	 
	We then used the linear regression algorithm separately on each sample group, obtaining a piecewise linear scoring function:}
	\item[Outcome:]  \hfill\\
	$$\sigma_{Xd}^{\textit{thi}} =
	\left\{ \begin{aligned}
	& 0.5681 \textit{thi} + 0.2194\textit{tyi} +\\
		&\enspace+0.1205 \textit{rat}+0.1492\textit{rch}+\\
		&\enspace\; - 1.9663 &
			\mbox{if $\textit{thi} \in [0,6)$;}\\
	& 0.3466 \textit{thi} + 0.4685\textit{tyi} +\\
		&\enspace+ 0.1476\textit{rat} + 0.2657\textit{rch}\\
		&\enspace\;+0.1567\textit{frn}-3.6959 
			&\mbox{if $\textit{thi} \in [6,8)$;}\\
	&0.7396\textit{thi} + 0.3527\textit{tyi} + \\
		&\enspace 0.1853  \textit{rch}+ 0.1017\textit{frn}+\\
		&\enspace\; - 4.1450 	
			&\mbox{if $\textit{thi} \in [8,10]$.}
	\end{aligned}\right.$$
	\end{description}	

\item \textbf{\icd/tyi} - use all five factors, dynamic approach depending on type interest.
	\begin{description}
	\item[Assumption:] {\em we assume that the relevance of additional factors depends on type interest. Similarly to what we did in the previous case, we split the samples in the training set into three groups of about 1000 samples, according to the value of factor {\em tyi} (type interest): a set of samples with $\textit{tyi} \in [0,6)$, a set of samples with $\textit{tyi} \in [6,8)$, and a set of samples with $\textit{tyi} \in [8,10]$.
Again, we obtain a piecewise linear scoring function:}
	\item[Outcome:]  \hfill\\
	$$\sigma_{Xd}^{\textit{tyi}} =
	\left\{ \begin{aligned}
	& 0.4410 \textit{thi} + 0.2786\textit{tyi} +\\
		&\enspace+0.1405\textit{rch}+0.1341\textit{frn}\\
		&\enspace\; - 1.5022 &
			\mbox{if $\textit{tyi} \in [0,6)$;}\\
	& 0.6467 \textit{thi} + 0.2798\textit{tyi} +\\
		&\enspace+ 0.1754\textit{rch}+0.0425\textit{frn}+\\
		&\enspace\;-2.3500 
			&\mbox{if $\textit{tyi} \in [6,8)$;}\\
	&0.6384\textit{thi} + 0.2181  \textit{rch}+\\
		&\enspace\; + 0.0823 	
			&\mbox{if $\textit{tyi} \in [8,10]$.}
	\end{aligned}\right.$$
	\end{description}		
\end{enumerate}		

\subsection{Testing}\label{sec:testing}
\begin{table*}[t]
\begin{center}
\begin{tabular}{lccccccc}
\toprule
 & {$\sigma^{\textit{init}}_0$} & {$\sigma^{\textit{fin}}_0$} & {$\sigma_X$} & {$\sigma^{\textit{abs}}_{Xu}$} & {$\sigma^{\textit{rel}}_{Xu}$} & {$\sigma_{Xd}^{\textit{thi}}$} & {$\sigma_{Xd}^{\textit{tyi}}$} \\ \otoprule
1 & 2.3032 & 2.6431 & 2.5793 & 2.6068 & 2.5926 & 2.5763 & 2.5412 \\ \midrule
2 & 2.3817 & 2.661 & 2.5916 & 2.6683 & 2.6307 & 2.5705 & 2.5747 \\ \midrule
3 & 2.4121 & 2.7205 & 2.6329 & 2.6877 & 2.67 & 2.6477 & 2.6337 \\ \midrule
4 & 2.5701 & 2.7754 & 2.7498 & 2.7605 & 2.7439 & 2.7438 & 2.7283 \\ \midrule
5 & 2.4279 & 2.6324 & 2.6094 & 2.6362 & 2.5964 & 2.6053 & 2.6094 \\ \midrule
6 & 2.3248 & 2.5663 & 2.5061 & 2.5513 & 2.5209 & 2.4852 & 2.477 \\ \midrule
7 & 2.4352 & 2.6901 & 2.6288 & 2.6632 & 2.6526 & 2.602 & 2.6028 \\ \midrule
8 & 2.3447 & 2.6944 & 2.6223 & 2.681 & 2.6496 & 2.6089 & 2.6043 \\ \midrule
9 & 2.3429 & 2.6541 & 2.5876 & 2.6421 & 2.6161 & 2.5674 & 2.5686 \\ \midrule
10 & 2.5097 & 2.7655 & 2.6819 & 2.7313 & 2.7078 & 2.6739 & 2.6637 \\ \midrule
11 & 2.4206 & 2.6379 & 2.561 & 2.6162 & 2.5901 & 2.5538 & 2.5334 \\ \midrule
12 & 2.4113 & 2.5892 & 2.5219 & 2.555 & 2.5347 & 2.5012 & 2.5022 \\ \midrule
13 & 2.4454 & 2.7071 & 2.665 & 2.6879 & 2.7048 & 2.6462 & 2.6469 \\ \midrule
14 & 2.3713 & 2.6402 & 2.5783 & 2.6372 & 2.6158 & 2.5506 & 2.5399 \\ \midrule
15 & 2.4711 & 2.6199 & 2.5665 & 2.5933 & 2.5915 & 2.5587 & 2.5752 \\ \bottomrule
\end{tabular}
  \caption{Validation results for the 15 random splits}\label{tab:results:val}
  \end{center}
\end{table*}
Each scoring function was trained 15 times, with a different random split of $S_{\textit{tr}}$ into a training and a validation set. The scoring functions were then tested within the Weka environment on the test set $S_{\textit{test}}$. 

\begin{table*}[t]
\begin{center}
\begin{tabular}{lcccccc}
\toprule
 & {$\sigma^{\textit{fin}}_0$} & {$\sigma_X$} & {$\sigma^{\textit{abs}}_{Xu}$} & {$\sigma^{\textit{rel}}_{Xu}$} & {$\sigma_{Xd}^{\textit{thi}}$} & {$\sigma_{Xd}^{\textit{tyi}}$} \\ \otoprule
1 & 2.585 & 2.482 & 2.549 & 2.5217 & 2.5058 & 2.4755 \\ \midrule
2 & 2.5885 & 2.4778 & 2.5522 & 2.519 & 2.4849 & 2.471 \\ \midrule
3 & 2.5943 & 2.4857 & 2.5549 & 2.5283 & 2.5269 & 2.4907 \\ \midrule
4 & 2.589 & 2.4841 & 2.5516 & 2.524 & 2.5103 & 2.4872 \\ \midrule
5 & 2.5879 & 2.4848 & 2.5524 & 2.521 & 2.4927 & 2.4854 \\ \midrule
6 & 2.5872 & 2.4784 & 2.549 & 2.5202 & 2.5004 & 2.4725 \\ \midrule
7 & 2.5897 & 2.4921 & 2.555 & 2.5269 & 2.491 & 2.4886 \\ \midrule
8 & 2.5908 & 2.4797 & 2.5506 & 2.5225 & 2.5002 & 2.4796 \\ \midrule
9 & 2.5838 & 2.4807 & 2.5512 & 2.5208 & 2.481 & 2.4846 \\ \midrule
10 & 2.5873 & 2.4829 & 2.552 & 2.5288 & 2.5047 & 2.4906 \\ \midrule
11 & 2.5866 & 2.479 & 2.5469 & 2.5198 & 2.5022 & 2.484 \\ \midrule
12 & 2.5901 & 2.4788 & 2.5487 & 2.5232 & 2.4832 & 2.4774 \\ \midrule
13 & 2.5875 & 2.4859 & 2.5486 & 2.5234 & 2.5016 & 2.4789 \\ \midrule
14 & 2.5942 & 2.4814 & 2.5578 & 2.5254 & 2.5073 & 2.4798 \\ \midrule
15 & 2.5893 & 2.4809 & 2.5498 & 2.5214 & 2.4996 & 2.4813 \\ \bottomrule
\end{tabular}
  \caption{Test results for the 15 random splits}
  \label{tab:results:test}
  \end{center}
\end{table*}

\begin{table}[b]
\begin{center}
\begin{tabular}{lccccc}
\toprule
 & {\textit{thi}} & {\textit{tyi}} & {\textit{rat}} & {\textit{rch}} & {\textit{frn}} \\ \otoprule
1 & 0.5761 & 0.3342 & 0.0997 & 0.1928 & 0.0902 \\ \midrule
2 & 0.567 & 0.3586 & 0.093 & 0.1867 & 0.1133 \\ \midrule
3 & 0.5879 & 0.3891 & 0.096 & 0.1726 & 0.1021 \\ \midrule
4 & 0.5891 & 0.3539 & 0.0717 & 0.2099 & 0.0851 \\ \midrule
5 & 0.5776 & 0.3531 & 0.105 & 0.2154 & 0.0949 \\ \midrule
6 & 0.5785 & 0.3478 & 0.0756 & 0.1901 & 0.0971 \\ \midrule
7 & 0.5883 & 0.3698 & 0.1208 & 0.1939 & 0.113 \\ \midrule
8 & 0.5719 & 0.3686 & 0.0846 & 0.1821 & 0.0882 \\ \midrule
9 & 0.5566 & 0.3256 & 0.0608 & 0.1847 & 0.0841 \\ \midrule
10 & 0.5754 & 0.356 & 0.356 & 0.1701 & 0.0792 \\ \midrule
11 & 0.5761 & 0.3483 & 0.0958 & 0.181 & 0.0849 \\ \midrule
12 & 0.5557 & 0.3661 & 0.1042 & 0.1912 & 0.0981 \\ \midrule
13 & 0.5897 & 0.3502 & 0.105 & 0.2022 & 0.1106 \\ \midrule
14 & 0.5952 & 0.3714 & 0.0596 & 0.1868 & 0.086 \\ \midrule
15 & 0.5708 & 0.3645 & 0.0886 & 0.1982 & 0.0938 \\ \bottomrule
\end{tabular}
\caption{Attribute coefficients for function $\sigma_X$} \label{tab:results:coeff}
  \end{center}
\end{table}

We collected the test results in terms of the RMSE (\textit{Root Mean Square Error}) between the predictions given by the scoring function (``system scores'') and the values explicitly provided by users (``user scores'') (see Formula~\ref{eq:RMSE}.). RMSE is the standard test measure for linear regression learning, and besides it is also a well-accepted statistical accuracy metric in the field of recommender systems \citep{Adomavicius:05:surv,Victor:11,Herlocker:04,Arazy:2010}.
A lower RMSE denotes a higher degree of accuracy; as a consequence, a negative variation in the RMSE is regarded as an improvement. As pointed out by Arazy et al.~\citep{Arazy:2010}: ``Even small RMSE improvements are considered valuable in the context of recommender systems. For example the Netflix prize competition\footnote{http://www.netflixprize.com} offered a one million dollar reward for an RMSE reduction of 10 percent."

Table ~\ref{tab:results:val} shows the results of validation, while table ~\ref{tab:results:test} shows the results of testing, for each of the random splits(numbered 1 through 15). For function $\sigma^{\textit{init}}_0$ we can provide only validation results, as the test samples did not include ``initial'' user scores and the function performance was therefore impossible to evaluate.

As we are also interested in analyzing the extent to which each factor contributes to the overall score, we show in table ~\ref{tab:results:coeff} the attributes coefficients of function $\sigma_X$ which, as we will see in the next section, has the lower RMSE values (i.e., it is the best-performing function).

\section{Discussion}
\label{sec:discussion}

Let us consider again the research questions we introduced in Section~\ref{sec:introduction} and discuss how the results of our experiments can provide us with answers.

Our first question concerns the consequences of not taking into account the additional factors in the recommendation process.
\begin{itemize}
\item [RQ1] What is the accuracy of pure content-based recommendation (i.e.\ considering only the themes and type) for events? As users typically know more about the event than its theme and type, how does this knowledge affect their choices and the recommendation accuracy? 
\end{itemize}

We attempt an answer to this question by checking the validity of hypothesis that {\em RMSE values for $\sigma^{\textit{init}}_0$ are worse (higher) than those for $\sigma^{\textit{fin}}_0$}.

According to results of the paired T-test, these RMSE values are significantly different, and the RMSE for $\sigma^{\textit{fin}}_0$ is on the average $10.6\%$ higher than the RMSE for $\sigma^{\textit{init}}_0$ with a $0.95$-confidence interval of $\pm1.2\%$.

In other words, if we ask users to evaluate their interest in participation based only on themes and type of the event, their answers are be quite different, and more similar to the score provided by the recommender. However, this is not the typical situation in real-life since users do know things such as the event location and its distance from their own location, they are aware of whether any of their friends would participate, and in a Web 2.0 context they are likely to know the opinion of the community about the event. Results show unmistakably that not taking the additional information into account is significantly detrimental for recommendation.

After evaluating the consequences of not considering the additional factors, our second research question investigates the advantages of including them. 
\begin{itemize}
\item [RQ2] If we provide the system with additional information, does this enhance recommender system accuracy? Namely, if we include reachability, average rating and friends participation in the recommendation, does the accuracy improve?
\end{itemize}

In this case we check the validity of the hypothesis that {\em RMSE values for $\sigma_X$ are better (lower) than those for $\sigma^{\textit{fin}}_0$}.

The paired T-test tells us that the hypothesis holds: RMSE for $\sigma_X$ is on the average $4.1\%$ lower than the RMSE for $\sigma^{\textit{fin}}_0$, with a $0.95$-confidence interval of $\pm0.1\%$.
Table \ref{tab:test:rq2} shows the paired T-test results for the comparison of $\sigma^{\textit{fin}}_0$ with all the other functions we trained which included additional factors. As it can be seen, they all improve over $\sigma^{\textit{fin}}_0$; best improvement is obtained however with with $\sigma_X$ and $\sigma_{Xd}^{\textit{tyi}}$.

\begin{table}[thc]
\begin{center}
\begin{tabular}{lccc}
\toprule
& $\Delta$ & $0.95$-conf. interval & t \\ \otoprule
$\sigma_X$ & 4.11\% & $\pm0.09\%$ & 95.392 \\ \midrule
$\sigma^{\textit{abs}}_{Xu}$ & 1.45\% & $\pm0.05\%$ & 59.567 \\ \midrule
$\sigma^{\textit{rel}}_{Xu}$ & 2.54\% & $\pm0.06\%$ & 87.645 \\ \midrule
$\sigma_{Xd}^{\textit{thi}}$ & 3.45\% & $\pm0.23\%$ & 32.269 \\ \midrule
$\sigma_{Xd}^{\textit{tyi}}$ & 4.13\% & $\pm0.14\%$ & 65.247 \\ \bottomrule
\end{tabular}
  \caption{Paired T-test comparison between $\sigma^{\textit{fin}}_0$ and other functions}
  \label{tab:test:rq2}
  \end{center}
\end{table}
Next, we investigate the relevance of each additional factor.

\begin{itemize}
\item [RQ3] 
To which extent does each additional factor contribute to the final score? 
\end{itemize}

\begin{table}[thc]
\begin{center}
\begin{tabular}{lccc}
\toprule
 & coefficient & {$0.95$-conf. interval} & {t} \\ \otoprule
\textit{thi} & 0.58 & $\pm0.0065$ & 190.524 \\ \midrule
\textit{tyi} & 0.36 & $\pm0.0087$ & 88.556 \\ \midrule
\textit{rat} & 0.090 & $\pm0.0096$ & 20.114 \\ \midrule
\textit{rch} & 0.19 & $\pm0.0069$ & 59.227 \\ \midrule
\textit{frn} & 0.094 & $\pm0.0061$ & 33.428 \\ \midrule
$R$ & 0.40 & $\pm-0.018$ & 47.59 \\ \bottomrule
\end{tabular}
  \caption{Average values and $0.95$-confidence intervals of $\sigma_{X}$ coefficients.}
  \label{tab:test:rq3}
  \end{center}
\end{table}
Table \ref{tab:test:rq3} shows the average values and the $0.95$-confidence intervals of coefficients learned for $\sigma_{X}$, which are listed in table \ref{tab:results:coeff}. It also shows the same information for the ratio $R$ between additional factors and content-based factors, which is about 4:10. It can be noted that {\em reachability} is the most relevant among additional factors, its weight being about twice that of {\em friend participation} and {\em ratings}.

An alternative option to set the coefficients is to directly ask the users about the importance of each factor for them. This brings us to the next research question:
\begin{itemize}
\item [RQ4] Would letting the users explicitly voice their preferences regarding additional factors provide an improvement in the recommendation? 
\end{itemize} 

In this case we check the validity of the hypothesis that {\em RMSE values for $\sigma_X$ are better (lower) than those for both $\sigma^{\textit{abs}}_{Xu}$ and $\sigma^{\textit{rel}}_{Xu}$.}

Table~\ref{tab:test:rq4} shows the comparison, by means of paired T-test, of these RMSE values. We can see  that both absolute and relative user-defined coefficients ($\sigma^{\textit{abs}}_{Xu}$ and $\sigma^{\textit{rel}}_{Xu}$) actually provide worse accuracy in recommendation than those calculated by linear regression ($\sigma_{X}$). 

\begin{table}[thc]
\begin{center}
\begin{tabular}{lccc}
\toprule
& $\Delta$ & $0.95$-conf. interval & {t} \\ \otoprule
$\sigma^{\textit{abs}}_{Xu}$ & -2.78\% & $\pm0.08\%$ & -74.666 \\ \midrule
$\sigma^{\textit{rel}}_{Xu}$ & -1.64\% & $\pm0.07\%$ & -52.224 \\ \bottomrule
\end{tabular}
  \caption{Paired T-test comparison between $\sigma_X$ and $\sigma_{Xu}$ functions.}
  \label{tab:test:rq4}
  \end{center}
\end{table}

Finally, we wanted to see whether the relevance of the additional factors actually depends on how much the user is interested in the event content. For example, distance or lack of friends may not matter much if a person is very interested, or if she is not interested at all, while they may count when the person is indecisive or has a moderate interest.
\begin{itemize}
\item [RQ5] Does the relevance of additional factors depend on how much the user is interested in the event content?
\end{itemize}

Here we can compare (Table \ref{tab:test:rq5}) the RMSE value for $\sigma_X$ with the RMSE  for $\sigma^{\textit{thi}}_{\textit{Xd}}$ and $\sigma^{\textit{tyi}}_{\textit{Xd}}$.

\begin{table}[thc]
\begin{center}
\begin{tabular}{lccc}
\toprule
& $\Delta$ & $0.95$-conf. interval & {t} \\ \otoprule
$\sigma^{\textit{thi}}_{Xd}$ & -0.69\% & $\pm0.25\%$ & -5.794 \\ \midrule
$\sigma^{\textit{tyi}}_{Xd}$ & 0.02\% & $\pm0.10\%$ & 0.383 \\ \bottomrule
\end{tabular}
  \caption{Paired T-test comparison between $\sigma_X$ and $\sigma_{Xd}$ functions.}
  \label{tab:test:rq5}
  \end{center}
\end{table}

We see that while $\sigma^{\textit{thi}}_{\textit{Xd}}$ is proven to actually behave {\em worse} than $\sigma_X$, $\sigma^{\textit{tyi}}_{\textit{Xd}}$ has the indicatively the same performance. The result is therefore not conclusive with respect to our investigation of RQ5: while the coefficients learned in the dynamic case (Table \ref{tab:test:rq5:coeff}) show that there is a relationship between the interest in the event content and the relevance of additional factors, neglecting this relationship and computing the coefficients over the whole set of samples does not cause a performance degradation.

\begin{table}[thc]
\begin{center}
\begin{tabular}{|l|l|l|r|r|r|}
\hline
 & attr. & range & \multicolumn{1}{c|}{coeff.} & \multicolumn{1}{c|}{$0.95$-conf.} & \multicolumn{1}{c|}{t} \\ \hline
\multicolumn{ 1}{|l|}{$\sigma^{\textit{thi}}_{\textit{Xd}}$} & \multicolumn{ 1}{l|}{\textit{rat}} & $[0,6)$ & .1037867 & $\pm.0247380$ & 8.998 \\ \cline{ 3- 6}
\multicolumn{ 1}{|l|}{} & \multicolumn{ 1}{l|}{} & $[6,8)$ & .1303267 & $\pm.0231980$ & 12.049 \\ \cline{ 3- 6}
\multicolumn{ 1}{|l|}{} & \multicolumn{ 1}{l|}{} & $[8,10]$ & 0 & $\pm 0$ & \multicolumn{1}{l|}{-} \\ \cline{ 2- 6}
\multicolumn{ 1}{|l|}{} & \multicolumn{ 1}{l|}{\textit{frn}} & $[0,6)$ & .0220600 & $\pm.0179181$ & 2.641 \\ \cline{ 3- 6}
\multicolumn{ 1}{|l|}{} & \multicolumn{ 1}{l|}{} & $[6,8)$ & .1477733 & $\pm.0122824$ & 25.805 \\ \cline{ 3- 6}
\multicolumn{ 1}{|l|}{} & \multicolumn{ 1}{l|}{} & $[8,10]$ & .1146600 & $\pm.0066689$ & 36.876 \\ \cline{ 2- 6}
\multicolumn{ 1}{|l|}{} & \multicolumn{ 1}{l|}{\textit{rch}} & $[0,6)$ & .1513533 & $\pm.0108975$ & 29.788 \\ \cline{ 3- 6}
\multicolumn{ 1}{|l|}{} & \multicolumn{ 1}{l|}{} & $[6,8)$ & .2379267 & $\pm.0137771$ & 37.040 \\ \cline{ 3- 6}
\multicolumn{ 1}{|l|}{} & \multicolumn{ 1}{l|}{} & $[8,10]$ & .1795067 & $\pm.0151628$ & 25.391 \\ \hline
\multicolumn{ 1}{|l|}{$\sigma^{\textit{tyi}}_{\textit{Xd}}$} & \multicolumn{ 1}{l|}{\textit{rat}} & $[0,6)$ & .0303200 & $\pm.0288319$ & 2.255 \\ \cline{ 3- 6}
\multicolumn{ 1}{|l|}{} & \multicolumn{ 1}{l|}{} & $[6,8)$ & .004867 & $\pm.0104380$ & 1.000 \\ \cline{ 3- 6}
\multicolumn{ 1}{|l|}{} & \multicolumn{ 1}{l|}{} & $[8,10]$ & .0215800 & $\pm.0248402$ & 1.863 \\ \cline{ 2- 6}
\multicolumn{ 1}{|l|}{} & \multicolumn{ 1}{l|}{\textit{frn}} & $[0,6)$ & .1441533 & $\pm.0134145$ & 23.048 \\ \cline{ 3- 6}
\multicolumn{ 1}{|l|}{} & \multicolumn{ 1}{l|}{} & $[6,8)$ & .0378800 & $\pm.0168071$ & 4.834 \\ \cline{ 3- 6}
\multicolumn{ 1}{|l|}{} & \multicolumn{ 1}{l|}{} & $[8,10]$ & .0225133 & $\pm.0170797$ & 2.827 \\ \cline{ 2- 6}
\multicolumn{ 1}{|l|}{} & \multicolumn{ 1}{l|}{\textit{rch}} & $[0,6)$ & .1457533 & $\pm.0113562$ & 27.528 \\ \cline{ 3- 6}
\multicolumn{ 1}{|l|}{} & \multicolumn{ 1}{l|}{} & $[6,8)$ & .1551733 & $\pm.0089711$ & 37.098 \\ \cline{ 3- 6}
\multicolumn{ 1}{|l|}{} & \multicolumn{ 1}{l|}{} & $[8,10]$ & .2178000 & $\pm.0129426$ & 36.093 \\ \hline
\end{tabular}
\caption{Average values and $0.95$-confidence intervals of $\sigma_{\textit{Xd}}$ coefficients.}  \label{tab:test:rq5:coeff}
  \end{center}
\end{table}

\section{Conclusions}
\label{sec:conclusions}

The goal of this work was to evaluate the impact of some social and contextual factors on user's interest in participating in an event, in order to evaluate their potential relevance in event recommendation. 
We started from the assumption that, in a social context, the interest a person has in an event depends not only on the type of activity that the event proposes, or the topic it concerns, but also on the reputation the event has in the community, on the participation of friends, and 
on the event location.
We then verified this assumption by experimental means, investigating how each of these factors (reputation, friends' participation and distance), together with basic information on the event content (type of activity and its theme or topic), can be taken into account in predicting the user interest, improving prediction accuracy.

In order to test our hypotheses we used linear regression to learn a predictive scoring function, expressed as a linear combination of the factors we wanted to study. Such scoring function can be seen as a form of content-based recommendation, since it tries to match event and user features in order to predict the user's interest. However, this does not imply that a recommender system exploiting such features should necessarily be a content-based one. Once the relevant information is represented in the user profile or object description, any recommendation technique can make use of it.

We use as a test-bed for experimenting our ideas a content-based recommender, which bases its suggestions on the features of the item to be recommended and on the profile of the user's interests. Hence, in this context it makes sense to consider the additional factors which influence the recommendation, since they can be easily accounted for in the user profiles and as features of the items to be recommended.

To the best of our knowledge, no other study in the literature on recommender systems tackles the same problem, although - as we discussed in Section \ref{sec:related} - the {\em type} of study is not new, as others (notably, \citep{Baltrunas:11},  \citep{Arazy:2010}) have investigated the impact of various factors in the recommendation of different items with a similar approach. 

Our study showed that the additional information provided by event reputation, friends' participation and distance actually influences users, as their explicit expressions of interest change quite significantly when they possess this information. Including this information in the scoring function indeed provided a good improvement on the score prediction.

Our user study also shows that asking users to self-evaluate how the additional factors influence their choices provides only a moderate improvement since the coefficients learned by linear regression for the whole population actually perform better than the personalized ones, when these are manually provided. 

This study has provided the basis for a prototype content-based recommender system described in ~\cite{Lombardi:13}, where the user can browse a list of events sorted by the prediction of a scoring function similar to the one we described, whose coefficients can however be manually tuned or completely switched off depending on the user's intention and context.
In this implementation, additional factors are further refined. In particular: ratings are weighted depending on the interest the raters have towards the theme or type of event; reachability is evaluated, together with time constraints, against the user's calendar; friend participation differentiates between users who are interested in the event and users who have confirmed the participation. 

The results presented in this paper are based on an analysis of predictions: we try to predict the scores that users would assign to events in order to state their interest. A different approach, very common in recommender systems, is to treat the recommendation as a classification problem, where the goal is to classify events as either recommendable or not, without ranking the items according to a score. Although, in principle, it is possible to see the scoring problem as a classification problem (each score is a class), the preliminary results we obtained with classification algorithms such as logistic regression were not satisfactory, with a very high RMSE. Indeed, with this type of algorithms RMSE is not a good measure of performance, since they do not take into account the ``distance'' between the actual class and the predicted one, but only the probability that the predicted class is correct. 
For this reason, we believe that a classification analysis would require a different study (possibly on the same data set), where the goal would be to determine whether an event should be recommended or not, rather than to predict a value expressing the user's interest.
  
Our user study focused on a large number of users, but a relatively small set of events. Other interesting analyses could be performed with a data set based on a large number of events, with users' interests expressed in a bigger time frame, possibly recording also {\em a posteriori} evaluation, after the event has been attended. This would allow us to learn a personalized combination of the scoring factors, as well as to measure time-related features. 

A larger range of events means also a larger range of event types and themes. While we maintain that our results prove that additional factors are worth including in a generic recommender system for events, and that in general users are not very good at assessing by themselves the importance of each factor, we also know that the extent to which each factor contributes to the overall recommendation is probably different for other types of events (for example, a person may go alone to an art exhibit with more ease than she would for a dinner).  

For this reason, we believe that  much can be learned from actually implementing a recommender for events, possibly as a Facebook application, in order to fully exploit social information as well as the event model already present within the popular social networking environment. This would allow having a much larger and more representative population, as well as a huge set of events and activities ranging from private parties to widespread social gatherings. Moreover, it can provide data for evaluating refined versions of the additional factors introduced here. The prototype described by \cite{Lombardi:13} is a first step in this direction.

\bibliographystyle{apalike}

\end{document}